\documentclass[a4paper,11pt]{article}
\pdfoutput=1 
\usepackage{jcappub} 
\usepackage[T1]{fontenc} 
\usepackage{amssymb,amsmath,amsfonts}
\usepackage{hyperref}  
\usepackage{float}
\usepackage{graphicx}
\usepackage{xcolor}
\usepackage{tensor}
\usepackage[utf8]{inputenc}
\usepackage{caption,subfig}
\usepackage{feynmf}

\title{Geometrized quantum Galileons}
\author[a]{Lavinia Heisenberg,}
\author[b]{Christian F. Steinwachs}
\affiliation[a]{Institute for Theoretical Physics, 
	ETH Zurich, Wolfgang-Pauli-Strasse 27, 8093, Zurich, Switzerland}
\affiliation[b]{Physikalisches Institut, Albert-Ludwigs-Universit\"at Freiburg,
	Hermann-Herder-Str.~3, 79104 Freiburg, Germany}
\emailAdd{lavinia.heisenberg@phys.ethz.ch}
\emailAdd{christian.steinwachs@physik.uni-freiburg.de}
\abstract{
	We investigate the renormalization structure of the scalar Galileon model in flat spacetime by calculating the one-loop divergences in a closed geometric form. The geometric formulation is based on the definition of an effective Galileon metric and allows to apply known heat-kernel techniques. The result for the one-loop divergences is compactly expressed in terms of curvature invariants of the effective Galileon metric and corresponds to a resummation of the divergent one-loop contributions of all $n$-point functions. The divergent part of the one-loop effective action therefore serves as generating functional for arbitrary $n$-point counterterms. We discuss our result within the Galileon effective field theory and give a brief outlook on extensions to more general Galileon models in curved spacetime.  
}

\begin{document}
	\allowdisplaybreaks[1]
	\maketitle
	\flushbottom
	%
	\section{Introduction}
	\label{Sec:Introduction}
	Many alternative theories of gravity invoke new dynamical degrees of freedom.
	In the geometrical framework these fields manifest themselves either through torsion or non-metricity, whereas in the field theoretical framework they appear as additional scalar, vector and tensor fields \cite{Heisenberg2018,Heisenberg2019a}.
	Generalized vector field models in cosmology have been investigated in \cite{Ford1989,Golovnev2008,Esposito-Farese2010,Tasinato2014,Heisenberg2014,BeltranJimenez2016,Allys2016,Heisenberg2018,Heisenberg2019a,DeFelice2016,DeFelice2016a}. The renormalization structure of the generalized Proca theory in curved spacetime has been discussed in \cite{Buchbinder2007,Toms2014,Toms2015,Belokogne2016, Buchbinder2017,Ruf2018b,Ruf2018a}.
	The simplest models with an additional propagating degree of freedom, are scalar-tensor theories and geometric modifications such as $f(R)$ gravity, see e.g. \cite{Sotiriou2010,DeFelice2010,Clifton2012,Nojiri2017}. The renormalization structure of these models on a general curved background in the one-loop approximation has been discussed in \cite{Barvinsky1993,Shapiro1995,Steinwachs2011,Steinwachs2012,Rham2013,Kamenshchik2015,Ruf2018,Ruf2018c}.
	
	A special class of effective field theories, the Galileon model, arises in the decoupling limit of the prominent theories of DGP \cite{Dvali2000}, massive gravity \cite{Rham2011} and generalized Proca \cite{Heisenberg2014,BeltranJimenez2016,Allys2016}.
	The inception of the non-linear interactions of the helicity-0 mode in the decoupling limit of the DGP model has motivated the construction of more general Galilean invariant interactions \cite{Nicolis2009,Deffayet2009a}.
	In the higher dimensional brane-world scenario the invariance under internal Galilean and shift transformations is just a reminiscent of the five dimensional Poincar\'e invariance.
	Independent of the specific embedding, from a four dimensional effective field theory perspective, one can attempt to build the most general Lagrangian for the Galileon scalar field, which gives rise to second-order equations of motion and is invariant under the Galilean transformation.
	The absence of higher order equations of motion ensures the absence of propagating ghost degrees of freedom.
	In four dimensional spacetime the construction allows only five non-trivial tree-level operators of the Galileon scalar field, which undergo a relative tuning among themselves in order to guarantee the second-order nature of the equations of motion.
	For the validity of the effective field theory, a detuning of the relative coefficients should not happen below the cutoff scale of the theory.
	While this is not the case for the scalar Galileon in flat spacetime, it becomes relevant in the generalization to curved space.
	
	Various aspects of the scalar Galileon at the quantum level have been studied previously in \cite{Nicolis2004,Hinterbichler2010,PaulaNetto2012,Rham2013,Heisenberg2014a,Brouzakis2014,Brouzakis2014a,Kampf2014,Heisenberg2014,Pirtskhalava2015,Saltas2017,Heisenberg2019}.
	The derivation of counterterms in the $\overline{\mathrm{MS}}$ scheme only requires the calculation of the ultraviolet (UV) divergent part for which there are very efficient specialized methods. 
	In \cite{PaulaNetto2012}, the one-loop effective action in position space was expanded up to quadratic powers of the Galileon field and the calculation reduced to a sum of universal functional traces introduced in the context of the generalized Schwinger-DeWitt formalism in \cite{Barvinsky1985}. The authors of \cite{PaulaNetto2012} also performed a complementary calculation by traditional Feynman diagrammatic methods, see e.g. \cite{Ellis2012} for a review on these Feynman diagrammatic methods. In, \cite{Kampf2014} the divergent part of the on-shell one-loop $4$-point amplitude (which includes the on-shell contributions to the divergent $1$PI one-loop diagrams) has been calculated by modern on-shell methods, which are in particular efficient for a high number of external legs, see e.g. \cite{Dixon2015,Cheung2018} for introductory reviews on these techniques.
	These results were generalized in \cite{Heisenberg2019}.
	Based on the implementation of the recently proposed combinatoric algorithm, specifically designed for the extraction of the UV divergent contributions of Feynman integrals for higher derivative theories \cite{Steinwachs2019}, the UV divergent contributions to the Galileon off-shell one-loop $n$-point correlation functions have been calculated in \cite{Heisenberg2019} up to $n=5$.
	
	In this paper we generalize these results to arbitrary $n$-point functions by a geometrical formulation.
	Beside diagrammatic momentum space methods, which are mainly used in the context of particle physics related calculations in flat spacetime, a very efficient approach to extract the one-loop divergences in curved spacetime is based on the combination of the background field method and the heat-kernel technique \cite{Schwinger1961,DeWitt1964,Atiyah1973,Gilkey1975,Abbott1982,Barvinsky1985,Avramidi2000,Vassilevich2003}. 
	The main strengths of this approach are its manifest covariance and its universality. Moreover, for quantum field theories with minimal second-order fluctuation operator, the Schwinger-DeWitt technique provides a closed algorithm for the calculation of the divergent part of the one-loop effective action \cite{DeWitt1965,Barvinsky1985}.
	In the geometric reformulation of this paper, we define an effective  Galileon metric constructed from derivatives of the scalar Galileon field. This effective metric, and the associated metric compatible connection define a generalized Laplacian in terms of which the Galileon fluctuation operator can be written as a minimal second-order operator.
	For this operator, the Schwinger-DeWitt method is directly applicable and the divergent part of the one-loop effective action is obtained in terms of geometrical invariants constructed from the effective Galileon metric and corresponds to a resummation of the divergent contributions of all $n$-point correlation functions. The divergent part of the one-loop effective action therefore serves as generating functional from which arbitrary $n$-point one-loop counterterms can be derived by successive functional differentiation.
	
	The article is structured as follows: In Sec.~\ref{model} we introduce the scalar Galileon model.
	In Sec.~\ref{1leffact} we derive the effective Galileon metric and its associated generalized Laplacian. We express the fluctuation operator of the Galileon action as a covariant minimal second-order operator and make use of the Schwinger-DeWitt technique to calculate the divergent part of the one-loop effective action in terms of geometric invariants constructed from the effective Galileon metric.
	In Sec.~\ref{genfunct} we discuss the role of the geometrically defined one-loop effective action as generating functional for arbitrary $n$-point counterterms.
	In Sec.~\ref{efffield}, we discuss how the geometrically defined effective action can be understood as a resummation of all one-loop $n$-point divergences in the context of the Galileon effective field theory. 
	We conclude in Sec.~\ref{conclusion} and give a brief outlook on possible generalization in curved spacetime.     
	
	Technical details are collected in several appendices. In App.~\ref{ExpInH}, we provide the expansion of the one-loop effective action up to fourth order in the perturbation of the inverse effective Galileon metric. Based on the results derived in App.~\ref{ExpInH}, we derive the expansion of the one-loop effective action up to fourth order in the Galileon field in App.~\ref{ExpInPi}. In App.~\ref{ccheck} we perform several crosschecks with results previously obtained in the literature. 
	
	\section{Galileon Model}
	\label{model}
	The action functional for the Galileon scalar field $\pi(x)$ in $d=4$ flat Euclidean space $\mathcal{M}$ with metric $g_{\mu\nu}=\delta_{\mu\nu}=\text{diag}(1,1,1,1)$ reads\footnote{We neglect tadpole contributions $\mathcal{L}_1=c_1M^3\pi$ and use the freedom in redefining (rescaling) the Galileon field and the couplings $c_i$ to choose $c_2=1/12$ such that the kinetic term is canonically normalized.}
	\begin{align}
	S[\pi]={}&\int_{\mathcal{M}}\mathrm{d}^{4}x\,\mathcal{L}(\pi,\partial\pi,\partial^2\pi),\qquad \mathcal{L}=\sum_{i=2}^{5}\mathcal{L}_{i},\label{Gact}\\
	\mathcal{L}_{2}={}&\frac{1}{12}\pi\epsilon^{\mu\nu\rho\sigma}\tensor{\epsilon}{^{\alpha}_{\nu\rho\sigma}}\pi_{\mu\alpha},\label{L1}\\
	\mathcal{L}_{3}={}&\frac{c_3}{M^3}\pi\epsilon^{\mu\nu\rho\sigma}\tensor{\epsilon}{^{\alpha\beta}_{\rho\sigma}}\pi_{\mu\alpha}\pi_{\nu\beta},\\
	\mathcal{L}_{4}={}&\frac{c_4}{M^6}\pi\epsilon^{\mu\nu\rho\sigma}\tensor{\epsilon}{^{\alpha\beta\gamma}_{\sigma}}\pi_{\mu\alpha}\pi_{\nu\beta}\pi_{\rho\gamma},\\
	\mathcal{L}_{5}={}&\frac{c_5}{M^9}\pi\epsilon^{\mu\nu\rho\sigma}\tensor{\epsilon}{^{\alpha\beta\gamma\delta}}\pi_{\mu\alpha}\pi_{\nu\beta}\pi_{\rho\gamma}\pi_{\sigma\delta}\label{L4}\,.
	\end{align}
	The second derivatives are defined as the symmetric tensor
	\begin{align}
	\pi_{\mu\nu}:=\partial_{\mu}\partial_{\nu}\pi.
	\end{align}
	The Galileon field and the partial derivatives have mass dimension $[\pi]=[\partial_{\mu}]=1$ (in natural units) and the operators \eqref{L1}-\eqref{L4} are expressed in units of the mass scale $M$ such that the coupling constants $c_i$ are dimensionless numbers.
	Since the Galileon action only involves derivative interactions, it is obviously invariant under shift symmetries $\pi\to\pi+c$ with a constant $c$. Modulo total derivatives, \eqref{Gact} is even invariant under the larger class of Galilei transformations with a constant vector $v_{\mu}$,
	\begin{align}
	\pi\to\pi+c+ v_{\mu}x^{\mu}.\label{Gsym}
	\end{align}
	This invariance is responsible for the particular structure of the Galileon interactions \eqref{L1}-\eqref{L4} and ensures that, despite the presence of higher derivative terms, the field equations are of second-order and no ghost-like excitations appear in the spectrum.
	
	\section{One-loop effective action}
	\label{1leffact}
	We split the Galileon field into background plus perturbation
	\begin{align}
	\pi(x)=\bar{\pi}(x)+\delta\pi(x).
	\end{align}
	The one-loop contribution to the Euclidean effective action is given by
	\begin{align}
	\Gamma_{1}[\langle\pi\rangle]=\frac{1}{2}\text{Tr}\ln F(\partial^{x})\delta(x,x')\label{OLG}.
	\end{align}
	The scalar second-order fluctuation operator $F(\partial)$ is defined by the Hessian 
	\begin{align}
	F(\partial^{x})\delta(x,x')=\left.\frac{\delta^2 S_{\mathrm{G}}[\pi]}{\delta\pi(x)\delta\pi(x')}\right|_{\pi=\bar{\pi}}=-\left(\bar{G}^{-1}\right)^{\mu\nu}\partial_{\mu}^{x}\partial_{\nu}^{x}\delta(x,x').\label{Op}
	\end{align}
	Within the one-loop approximation \eqref{OLG}, the mean field $\langle\pi\rangle$ might be identified with the background field $\bar{\pi}$. In what follows we omit the bar indicating a background quantity.
	As indicated by the superscript, the derivatives $\partial^x$ in \eqref{Op} act on the first argument $x$ of the delta function. The symmetric tensor $\left(G^{-1}\right)^{\mu\nu}$ is defined in terms of the Galileon field by
	\begin{align}
	\left(G^{-1}\right)^{\mu\nu}:=-{}&\left(\frac{1}{6}\varepsilon^{\mu\alpha\rho\sigma}
	{{\varepsilon}^{\nu}}_{\alpha\rho\sigma}+\frac{6}{M^3}{\varepsilon}^{\mu\alpha\rho\sigma}
	{{\varepsilon}^{\nu\beta}}_{\rho\sigma}\pi_{\alpha\beta} \right.\nonumber\\
	&\left.+12\frac{c_4}{M^6}{\varepsilon}^{\mu\alpha\rho\sigma}
	{{\varepsilon}^{\nu\beta\gamma}}_{\sigma}\pi_{\alpha\beta}\pi_{\rho\gamma}+20\frac{c_5}{M^9}{\varepsilon}^{\mu\alpha\rho\sigma}
	{{\varepsilon}^{\nu\beta\gamma\kappa}}\pi_{\alpha\beta}\pi_{\rho\gamma}\pi_{\sigma\kappa}\right)\,.\label{GMetric}
	\end{align}
	The structure of the operator \eqref{Op} naturally suggests to identify $\left(G^{-1}\right)^{\mu\nu}$ with the inverse of an ``effective Galileon metric'' $G_{\mu\nu}$, defined in terms of \eqref{GMetric} via 
	\begin{align}
	G_{\mu\rho}\left(G^{-1}\right)^{\rho\nu}=\delta^{\nu}_{\mu}.
	\end{align}
	The Galileon metric $G_{\mu\nu}$ is assumed to be positive definite to ensure its non-degeneracy\footnote{Positive definiteness requires all eigenvalues of $G_{\mu\nu}$ to be positive. Since the eigenvalues are functions of the $c_i$, the condition of positive definiteness implies constraints on the $c_i$.}
	\begin{align}
	\det(G)\neq 0.
	\end{align}
	While $G_{\mu\nu}$ is not compatible with $\partial_{\mu}$, the particular structure of \eqref{GMetric} leads to the important property that $\left(G^{-1}\right)^{\mu\nu}$ is divergence-free 
	\begin{align}
	\partial_{\mu}\left(G^{-1}\right)^{\mu\nu}=0.\label{GinvDfree}
	\end{align}
	The effective Galileon metric \eqref{GMetric} defines the required geometric structure for an application of the covariant heat-kernel techniques for the fluctuation operator \eqref{Op}.  
	
	\subsection{Covariant reformulation in terms of a minimal second-order operator }
	We define $\nabla^{G}_{\mu}$ to be the torsion-free covariant derivative compatible with $G_{\mu\nu}$,
	\begin{align}
	[\nabla^{G}_{\mu},\nabla^{G}_{\nu}]\phi=0,\qquad
	\nabla^{G}_{\rho}G_{\mu\nu}=0,
	\end{align}
	for some scalar field $\phi(x)$. Clearly the connection $\tensor{\Gamma}{^{\rho}_{\mu\nu}}(G)$ associated to $\nabla^{G}$ reads
	\begin{align}
	\tensor{\Gamma}{^{\rho}_{\mu\nu}}(G)=\frac{\left(G^{-1}\right)^{\rho\sigma}}{2}\left(\partial_{\mu}G_{\sigma\nu}+\partial_{\nu}G_{\sigma\mu}-\partial_{\sigma}G_{\mu\nu}\right).\label{GGam}
	\end{align}
	From now on, we lower and raise indices exclusively with $G_{\mu\nu}$ and $\left(G^{-1}\right)^{\mu\nu}$, respectively. We define the positive definite  covariant Laplacian as
	\begin{align}
	\Delta_{G}:=-\left(G^{-1}\right)^{\mu\nu}\nabla^{G}_{\mu}\nabla^{G}_{\nu}.\label{GLap}
	\end{align}
	When the Laplacian $\Delta_{G}$ acts on scalars, it is related to the fluctuation operator \eqref{Op} by
	\begin{align}
	F(\partial)=-\left(G^{-1}\right)^{\mu\nu}\partial_{\mu}\partial_{\nu}=\Delta_{G}-\left(G^{-1}\right)^{\mu\nu}\tensor{\Gamma}{^{\rho}_{\mu\nu}}(G)\nabla^{G}_{\rho}.\label{LapG}
	\end{align}
	In addition, we define the ``bundle connection'' acting on scalars
	\begin{align}
	\Sigma^{\rho}:=\frac{1}{2}\left(G^{-1}\right)^{\mu\nu}\tensor{\Gamma}{^{\rho}_{\mu\nu}}.\label{GCon}
	\end{align} 
	Combining \eqref{LapG} with \eqref{GCon}, the operator \eqref{GLap} can be expressed in the covariant form
	\begin{align}
	F(\nabla^{G})=\Delta_{G}-2\Sigma^{\rho}\nabla_{\rho}^{G}.\label{Op2}
	\end{align}
	In terms of $\mathcal{D}_{\mu}:=\nabla^{G}_{\mu}+G_{\mu\nu}\Sigma^{\nu}$, the operator \eqref{Op2} acquires the minimal second-order form
	\begin{align}
	F(\mathcal{D})=-\mathcal{D}_{\mu}\mathcal{D}^{\mu}+P.\label{Op3}
	\end{align}
	The terms resulting from the redefinition $\nabla^{G}\to\mathcal{D}$ are absorbed in the potential part $P$,
	\begin{align}
	P:=\nabla_{\nu}^{G}\Sigma^{\nu}+\Sigma_{\nu}\Sigma^{\nu}.\label{PotP}
	\end{align}
	The bundle curvature $\mathcal{R}_{\mu\nu}$ vanishes due to the antisymmetry and the scalar nature of $\pi$,
	\begin{align}
	\mathcal{R}_{\mu\nu}\pi:=[\mathcal{D}_{\mu},\mathcal{D}_{\nu}]\pi=0.
	\end{align}
	
	\subsection{Final result of the one-loop divergences in a closed form}
	For the minimal second-order operator \eqref{Op3}, the Schwinger-DeWitt algorithm is directly applicable and the one-loop divergences are expressed in a closed form in terms of the quadratic curvature invariants of the effective Galileon metric $G_{\mu\nu}$ and the potential $P$,
	\begin{align}
	\Gamma_{1}^{\mathrm{div}}[G]={}&-\frac{\Lambda}{2}\int_{\mathcal{M}}\mathrm{d}^4x\det(G)^{1/2}\left\{\frac{1}{60}\left(G^{-1}\right)^{\mu\rho}\left( G^{-1}\right)^{\nu\sigma}R_{\mu\nu}(G)R_{\rho\sigma}(G)\right.\nonumber\\
	&\phantom{-\frac{\chi(\mathcal{M})}{180}}\left.+\frac{1}{120}R^2(G)-\frac{1}{6}R(G)P(G)+\frac{1}{2}P^2(G)\right\}-\frac{\chi(\mathcal{M)}}{180\varepsilon}.\label{OneLoopG}
	\end{align}
	In \eqref{OneLoopG} we have absorbed the pole in dimension $1/\varepsilon$ with the factor $1/(4\pi)^2$ in the definition
	\begin{align}
	\Lambda:=\frac{1}{(4\pi)^2\varepsilon}.
	\end{align}
	The Euler characteristic $\chi(\mathcal{M})$ of the manifold $\mathcal{M}$ in $d=4$ dimensions is a topological term, independent of the metric $G_{\mu\nu}$, and defined in terms of the Gauss-Bonnet invariant $\mathcal{G}$,
	\begin{align}
	\chi(\mathcal{M}):=&{}\frac{1}{32\pi^2}\int_{\mathcal{M}}\mathrm{d}^4x\det(G)^{1/2}\mathcal{G}(G),\\ \mathcal{G}(G):=&{}R_{\mu\nu\rho\sigma}(G)R^{\mu\nu\rho\sigma}(G)-4R_{\mu\nu}(G)R^{\mu\nu}(G)+R^2(G)\,.
	\end{align}
	The one-loop divergences \eqref{OneLoopG} for the scalar Galileon \eqref{Gact} expressed in terms of curvature invariants of the effective Galileon metric \eqref{GMetric} constitutes our main result. It corresponds to a particular resummation of all $n$-point off-shell one-loop divergences and expressed in terms of curvature invariants of the effective Galileon metric. This generalizes previous results, obtained for the divergent part of the off-shell $2$-point function \cite{PaulaNetto2012}, the on-shell $4$-point function \cite{Kampf2014} and the off-shell $n$-point functions for $n=1,\ldots,5$ \cite{Heisenberg2019}. As discussed in more detail in Sec.~\ref{genfunct}, the results for a given $n$-point function can be recovered from the geometrically defined one-loop effective action by $n$-fold  functional differentiation. Hence, \eqref{OneLoopG} serves as generating functional for all one-loop counterterms. 
	
	Several non-trivial checks of \eqref{OneLoopG} are provided in App.~\ref{ccheck} by comparing the results for $n$-point functions, obtained from a systematic expansion of the generating functional \eqref{OneLoopG}, with the results obtained by direct Feynman diagrammatic calculations \cite{PaulaNetto2012,Kampf2014, Heisenberg2019}.

	\section{Generating functional of one-loop $n$-point counterterms}
	\label{genfunct}
	The result for the divergent part of the one-loop effective action \eqref{OneLoopG} serves as generating functional for the one-loop counterterms of arbitrary high $n$-point correlation functions
	\begin{align}
	\left.\langle\pi(x_1)\ldots \pi(x_n)\rangle^{\mathrm{div}}:=\frac{\delta^n\Gamma_{1}^{\mathrm{div}}[\pi]}{\delta \pi(x_1)\ldots\delta\pi(x_n)}\right|_{\pi=0}.\label{npoint}
	\end{align} 
	We first expand \eqref{OneLoopG} up to the $n$th power of the perturbations of the inverse metric\footnote{Instead of performing the expansion of \eqref{OneLoopG} via the ``chain-rule'', by using first expanding in perturbations of the inverse metric \eqref{Hexp} and subsequently express the inverse metric perturbations in terms of the perturbations of the Galileon field $\pi$, the geometric invariants in \eqref{OneLoopG} could be first expressed in terms of the Galileon field and the functional derivatives taken with respect to $\pi$. The explicit expression for the geometric invariants in terms of derivatives of the Galileon field can e.g. be obtained via the Cayley-Hamilton theorem.}
	\begin{align}
	\left(G^{-1}\right)^{\mu\nu}=\left(\bar{G}^{-1}\right)^{\mu\nu}+\sum_{k=1}^{n}\xi^k H_{k}^{\mu\nu}.\label{Hexp}
	\end{align}
	The inverse ``background metric'' $\left(\bar{G}^{-1}\right)^{\mu\nu}$ is defined by \eqref{GMetric} for vanishing mean field $\pi=0$,
	\begin{align}
	\left(\bar{G}^{-1}\right)^{\mu\nu}:=\left(G^{-1}\right)^{\mu\nu}|_{\pi=0}=-\delta^{\mu\nu}, \qquad\bar{G}_{\mu\nu}:=G_{\mu\nu}|_{\pi=0}=-\delta_{\mu\nu}.\label{Gbar}
	\end{align} 
	The one-loop divergences \eqref{OneLoopG} expanded up to $\mathcal{O}\left(\xi^{n}\right)$ are given by the series
	\begin{align}
	\Gamma^{\mathrm{div}}_{1}=\sum_{k=0}^{n}\left.\xi^{k}\Gamma^{\mathrm{div}}_{1,k}\right|_{\xi=1},\label{ExpEffAct}
	\end{align}
	where $\Gamma^{\mathrm{div}}_{1,k}$ is the $k$-th variation of \eqref{OneLoopG} with respect to $\left(G^{-1}\right)^{\mu\nu}$ around $\left(\bar{G}^{-1}\right)^{\mu\nu}$,
	\begin{align}
	\Gamma^{\mathrm{div}}_{1,k}:=\frac{1}{k!}\left.\delta^{k}_{G^{-1}}\Gamma^{\mathrm{div}}_{1}\right|_{G^{-1}=\bar{G}^{-1}}.
	\end{align}
	The $k$-th perturbation $H_{k}^{\mu\nu}$ is expressed in terms of the linear perturbations $\delta\pi$ by \eqref{GMetric},
	\begin{align}
	H_{k}^{\mu\nu}:=\left.\delta^{k}_{\pi}\left(G^{-1}\right)^{\mu\nu}\right|_{\pi=0},\qquad H_{k}:= H_{k}^{\mu\nu}\bar{G}_{\mu\nu},\quad k\geq 1.
	\end{align}
	The perturbations $H_{k}^{\mu\nu}$ inherit the important property \eqref{GinvDfree} of $\tensor{\left(G^{-1}\right)}{^{\mu\nu}}$,
	\begin{align}
	\partial_{\mu}H_{k}^{\mu\nu}=0.\label{DivFreeH}
	\end{align}
	Since the divergent part of the one-loop effective action \eqref{OneLoopG} is quadratic in curvatures, the zeroth and first order of the expansion $\Gamma^{\mathrm{div}}_{1,0}$ and $\Gamma^{\mathrm{div}}_{1,1}$ vanish for $\pi=0$ (corresponding to a flat Galileon background metric). For the same reason, only perturbations up to $H_{k-1}^{\mu\nu}$ enter the expansion $\Gamma^{\mathrm{div}}_{1,k}$.
	According to \eqref{GMetric}, $\tensor{\left(G^{-1}\right)}{^{\mu\nu}}$ is a third-order polynomial in $\pi$. Consequently, the $H_{k}^{\mu\nu}$ are vanishing for $k\geq4$. The explicit expressions for the non-vanishing $H_{k}^{\mu\nu}$ in terms of $\delta\pi$ read
	\begin{align}
	H_{1}^{\mu\nu}={}&\frac{12c_3}{M^3}\Big[\delta^{\mu\nu}\left(-\partial^2\delta\pi\right)+\left(\partial^{\mu}\partial^{\nu}\delta\pi\right)\Big],\label{H1}\\
	H_{2}^{\mu\nu}={}&\frac{24c_4}{M^6}\Big[-2\left(\partial^{\mu}\partial^{\rho}\delta\pi\right)\left(\partial^{\nu}\partial_{\rho}\delta \pi\right)-2\left(\partial^{\mu}\partial^{\nu}\delta \pi\right) \left(-\partial^2 \delta \pi\right)\nonumber\\
	&\qquad\;-\delta^{\mu\nu}\left(-\partial^2 \delta \pi\right)^2+\delta^{\mu\nu}\left(\partial_{\rho}\partial_{\sigma}\delta \pi\right)\left(\partial^{\rho}\partial^{\sigma}\delta\pi\right)\Big]\label{H2},\\
	H_{3}^{\mu\nu}={}&\frac{120c_5}{M^9}\Big[6\left(\partial^{\mu}\partial^{\rho}\delta\pi\right)\left(\partial^{\nu}\partial^{\sigma}\delta\pi\right)\left(\partial_{\rho}\partial_{\sigma}\delta\pi\right)+6\left(\partial^{\mu}\partial^{\rho}\delta\pi\right)\left(\partial^{\nu}\partial_{\rho}\delta\pi\right)\left(-\partial^2\delta\pi\right)\nonumber\\
	&\qquad\;\;+3\left(\partial^{\mu}\partial^{\nu}\delta\pi\right)\left(-\partial^2\delta\pi\right)^2-3\left(\partial^{\mu}\partial^{\nu}\delta\pi\right)\left(\partial^{\rho}\partial^{\sigma}\delta\pi\right)\left(\partial_{\rho}\partial_{\sigma}\delta\pi\right)\nonumber\\
	&\qquad\;\;+\delta^{\mu\nu}\left(-\partial^2 \delta\pi\right)^3-2\delta^{\mu\nu}\left(\partial^{\rho}\partial_{\sigma}\delta\pi\right)\left(\partial^{\sigma}\partial_{\lambda}\delta\pi\right)\left(\partial^{\lambda}\partial_{\rho}\delta\pi\right)\nonumber\\
	&\qquad\;\;-3\delta^{\mu\nu}\left(-\partial^2 \delta\pi\right)\left(\partial_{\rho}\partial_{\sigma}\delta\pi\right)(\partial^{\rho}\partial^{\sigma}\delta\pi)\Big].\label{H3}
	\end{align}
	Indices in \eqref{H1}-\eqref{H3} are raised and lowered with $\delta_{\mu\nu}$ and  $-\partial^2:=-\delta^{\mu\nu}\partial_{\mu}\partial_{\nu}$ defines the positive definite Laplacian. The one-loop counterterms for a given $n$-point function in terms of the Galileon field $\pi$ are obtained from the geometric result \eqref{OneLoopG} by inserting \eqref{H1}-\eqref{H3} in the expansion \eqref{ExpEffAct}.
	We explicitly demonstrate the results of the expansion \eqref{ExpEffAct} up to fourth order in the $H_{k}^{\mu\nu}$ in Appendix \ref{ExpInH} and provide the off-shell one-loop $n$-point function up to $n=4$, i.e. up to fourth order in $\pi$ in Appendix \ref{ExpInPi}.

	\section{Renormalization and Galileon effective field theory}
	\label{efffield}
	The Galileon theory \eqref{Gact} is perturbatively non-renormalizable and hence has to be considered as effective field theory (EFT).
	It is clear that the shift symmetry $\pi\to\pi+c$ prevents any monomial interactions $\pi^n/M^{n-4}$ from being radiatively generated -- only derivative interactions $\partial^n\pi^m/M^{n+m-4}$ are generated. In particular, already the first loop corrections only induce operators with at least second derivatives per field. These terms automatically satisfy the Galileon symmetry \eqref{Gsym}. Moreover, since these operators carry higher derivatives per field than the tree-level operators in the defining Galileon action \eqref{Gact}, it is clear that the operators in \eqref{L1}-\eqref{L4} are not renormalized. Nevertheless, the consistent renormalization of the Galileon effective field theory requires to take these higher derivative operators into account in a systematic way.
	
	As noted in \cite{Nicolis2004} and later in \cite{ Brouzakis2014,Brouzakis2014a}, the general structure of the divergent part of the one-loop effective action in $d=4$ has the schematic form (suppressing the index structure)
	\begin{align}
	\Gamma_{1}^{\mathrm{div}}=\int\mathrm{d}^4x\sum_k\left[M^4+M^2\partial^2+\partial^4\log\left(\frac{\partial^2}{M^2}\right)\right]\left(\frac{\partial^2\pi}{M^3}\right)^k.\label{EffActStruc}
	\end{align}
	In dimensional regularization only the last term in \eqref{EffActStruc} survives.\footnote{Dimensional regularization annihilates all power-law divergences and is only sensitive to the logarithmic divergences. Nevertheless, power law divergences which would arise in a different regularization scheme within the one-loop approximation of the Galileon in $d=4$, might still be calculate in dimensional regularization by ``dimensional reduction'', i.e.~quadratic divergences in $d=4$ are formally related to logarithmic divergences in $d=2$ and quartic divergences in $d=4$ to logarithmic divergences in $d=0$. In terms of the heat-kernel, logarithmically divergent contributions in lower dimensions are proportional to the integrated trace of the coincidence limit of the lower order Schwinger-DeWitt coefficient. Thus, in the case of the Galileon action \eqref{Gact} with effective Galileon metric \eqref{GMetric}, these divergences should again be expressible in a resummed way in terms of geometric invariants proportional to $\int\mathrm{d}^4x\,\mathrm{det}(G)^{1/2}\left[P(G)-R(G)/6\right]$ and $\int\mathrm{d}^4x\,\mathrm{det}(G)^{1/2}$, respectively.} 
	Similar to the discussion in \cite{Nicolis2004, Heisenberg2019} and by inspection of the structure \eqref{EffActStruc}, there are two dimensionless parameters which control the hierarchy among different operators in the Galileon effective field theory expansion
	\begin{align}
	\sigma_{\partial^2}:=\frac{\partial^2}{M^2},\qquad \sigma_{\partial^2\pi}:=\frac{\partial^2\pi}{M^3}.
	\end{align} 
	The ``classical'' parameter $\sigma_{\partial^2\pi}$ is related to the powers of derivatives per fields, i.e. for a fixed power of $\sigma_{\partial^2}$, the parameter $\sigma_{\partial^2\pi}$ counts the non-linearity of the theory, while the ``quantum'' parameter $\sigma_{\partial^2}$ is related to the number of derivatives and hence, for a fixed power of $\sigma_{\partial^2\pi}$, might be associated with the loop order.
	In fact, for phenomenological reliability of the Galileon model classical solutions with $\partial^2\pi/M^3\sim1$ exist, while quantum corrections are still under control $\partial^2/M^2\ll1$ \cite{Nicolis2004}.
	In terms of this classification, the one-loop result \eqref{OneLoopG}, expressed in terms of curvature invariants of the effective Galileon metric \eqref{GMetric}, corresponds to a geometric resummation of operators with an arbitrary number of fields, but a fixed number of derivatives per fields, i.e. arbitrary powers of $\sigma_{\partial^2\pi}$, but fixed powers of $\sigma_{\partial^2}$.
	In the absence of a UV completion or any heavy massive degree of freedom of a more fundamental theory, which, when integrated out, would set a natural cutoff scale for the validity of the resulting low energy effective Galileon theory, the only natural cutoff scale $\Lambda$ in the EFT expansion is the a priori unknown mass scale $M$ entering the Galileon action \eqref{Gact}. 
	
	Compared to the standard effective field theory expansion, the Galileon effective field theory is organized in a rather unusual way. Neither the naive expansion in inverse powers of the mass scale $M$, nor the expansion in powers of derivatives $\partial$, nor the expansion in powers of the field $\pi$ provide a correct expansion scheme according to which the higher derivative operators in the Galileon effective field theory are ordered. 
	The relevant parameter which organizes the Galileon EFT expansion is given by
	\begin{align}
	\sigma_{G}:=\#_{\pi}- 1/2\#_{ \partial}
	\end{align}
	and is determined by the structure of the corresponding operators with the number of $\pi$ fields $\#_{\pi}$ and the number of derivatives $\#_{\partial}$. Within this expansion scheme, all tree-level derivative interactions \eqref{L1}-\eqref{L4}, which define the low-energy limit of the Galileon EFT and hence the propagating degrees of freedom, are on equal footing, i.e. have $\sigma_G=1$.\footnote{The tree-level operators \eqref{L1}-\eqref{L4} are the lowest dimensional operators which satisfy the defining Galileon symmetry \eqref{Gsym} in a non-trivial way. This implies that the low-energy limit of the Galileon EFT is defined by all kinetic operators $\mathcal{L}_{2}-\mathcal{L}_{5}$ and not by the standard kinetic term $\mathcal{L}_{2}$ with the higher derivative terms $\mathcal{L}_{3}-\mathcal{L}_{5}$ treated as perturbations. Otherwise, there would already be a hierarchy among the operators in \eqref{Gact}.} Moreover, the higher dimensional operators induced by the one-loop divergences all have a homogeneous power $\sigma_G=-2$. In particular, all $n$-point operators which arise form the expansion of the geometrically resummed invariants when expressed in terms of the Galileon field $\pi$ have this property, i.e. share the same $\sigma_{G}$. By construction, this follows from the homogeneous $\sigma_G=0$ scaling of the terms which define the effective Galileon metric \eqref{GMetric} and the fact that the resummed one-loop divergences \eqref{OneLoopG} only involve terms proportional to curvatures squared (the potential $P$ also counts as curvature) and hence involve four additional derivatives compared to the tree-level interactions, in agreement with the structure of the logarithmic divergences in \eqref{EffActStruc}. Thus, this counting scheme is consistent with the geometric resummation, i.e.~allowing for arbitrary powers of $\sigma_{\partial^2\pi}$ but restrict to a fixed power of $\sigma_{\partial^2}$ consistent with the one-loop approximation.
	
	Although different in nature, the geometric resummation discussed here in the context of the Galileon theory shares some similarities to the situation in General Relativity (GR) in the following sense:
	In GR, starting from the linearized theory for a spin-2 particle propagating on flat spacetime, without knowledge of the full non-linear theory, symmetry (diffeomorphisms) dictates how to consistently add non-linear self-interactions in an iterative way. Resummation of these non-linearities into curvatures (which are invariant under infinitesimal diffeomorphisms) recovers the full, non-linear theory of GR along with its geometric interpretation \cite{Deser1970}. Since the linear theory is a second-order derivative theory, also all non-linear self-interactions only include up to two derivatives (a cosmological constant resummes into $\sqrt{- g})$. Since, each curvature comes with two derivatives, the resummed theory (GR) must be linear in curvature. Since the only linear curvature invariant is the Ricci scalar, this procedure uniquely results into the Einstein-Hilbert operator $\sqrt{-g}R$. This resummation of non-linearities is purely classical and leads to the exact full non-linear classical theory of GR. It has nothing to do with the EFT description of gravity, in which classically marginal and irrelevant higher dimensional operators (which are of course also constraint by diffeomorphism invariance and locality and have the form of scalar invariants constructed from powers of curvatures and derivatives thereof) are added to the Einstein-Hilbert term and treated as perturbations. Since GR is perturbatively non-renormalizable, this implies that at each order in the perturbative expansion new counterterms are required to cancel the ultraviolet divergences. It can be shown that these counterterms also respect the underlying diffeomorphism invariance and have the structure of local curvature invariants, see e.g. \cite{Barnich1994,Barnich1995,Barvinsky2018}.
	Despite the formal similarity of the geometric resummation of non-linear interaction terms, the geometric resummation of the one-loop divergences in the Galileon \eqref{OneLoopG} has a different origin and results from the particular structure of the Galileon fluctuation operator \eqref{Op} which suggests a covariant reformulation in terms of the effective Galileon metric \eqref{GMetric}.
	Thus, the only analogy of the resummation of the one-loop UV divergences in the Galileon and the (classical) resummation in GR is that both resum into curvatures. A priori, the resummation  of the Galileon one-loop divergences could have been very different, i.e. without knowledge of the effective geometric structure defined by the Galileon metric \eqref{GMetric}, the particular structure of the one-loop divergences in terms of quadratic curvature invariants would not have been obvious.

	Instead of a resummation of operators with arbitrary powers of $\sigma_{\partial^2\pi}$ and fixed order of $\sigma_{\partial^2}$, the opposite resummation with a fixed number of fields but an arbitrary number of derivatives might also be possible. As e.g. for the covariant perturbation theory discussed in \cite{Barvinsky1987,Barvinsky1990}, the resummation with a fixed number of curvatures but an infinite number of derivatives gives access to the non-local terms of the effective action, which, for a fixed order in the curvature expansion, can be represented in terms of non-local form factors. It would be interesting to study such a resummation in the context of the Galileon theory.

	\section{Conclusions}
	\label{conclusion}
	We have calculated the one-loop divergences for the scalar Galileon in flat spacetime. We obtained the result in a closed form in terms of curvature invariants of an effective Galileon metric, which appears naturally in the fluctuation operator of the Galileon. The effective Galileon metric defines a metric-compatible connection and a covariant Laplacian in terms of which the Galileon fluctuation operator can be written as second-order minimal operator. For such operators the Schwinger-DeWitt algorithm, which is based on a combination of the background field method and heat-kernel techniques, provides a closed algorithm for the calculation of the one-loop divergences. Consequently, the result for the one-loop divergences is expressed in terms of quadratic curvature invariants of the effective Galileon metric. The divergent part of the geometrically defined one-loop effective action serves as generating functional for the divergent part of all $n$-point correlation functions
	and corresponds to a resummation of the divergent contributions of all $n$-point functions.
	Therefore, our result \eqref{OneLoopG} generalizes previous calculations \cite{PaulaNetto2012,Kampf2014,Heisenberg2019} for the first few $n$-point functions to arbitrary $n$-point functions.
	
	We have demonstrated explicitly that for a given $n$, the divergent part of the $n$-point correlation function can be obtained from the divergent part of the geometrically defined one-loop effective action by $n$-fold functional differentiation. We performed this expansion up to $n=4$ and compared the resulting expressions with results obtained by Feynman diagrammatic momentum space methods \cite{PaulaNetto2012,Kampf2014,Heisenberg2019} . We found perfect agreement. This provides an independent check of our result as well as of the method based on the geometrical reformulation.  
	
	We also discussed the geometrical resummation in the Galileon effective field theory framework. It would be interesting to extend the Galileon effective field theory to curved spacetime \cite{Deffayet2009a}. In particular, it would be interesting to classify the possible structure of the counterterms, which, in view of the geometric resummation \eqref{OneLoopG} obtained in flat spacetime, would suggest that the one-loop divergences should be expressible in terms of scalar contractions among (derivatives of) curvatures of the spacetime metric $g_{\mu\nu}$ and the effective Galileon metric $G_{\mu\nu}$. 
	

	\acknowledgments
	L.H. is grateful to Achillefs Lazopoulos and Johannes Noller for useful discussions. C.S. thanks Michael Ruf for fruitful discussions. L.H. is supported by funding from the European Research Council (ERC) under the European Unions Horizon 2020 research and innovation programme grant agreement No 801781 and by the Swiss National Science Foundation grant 179740.

	\appendix
	
	
	\section{Expansion: Inverse metric perturbations}
	\label{ExpInH}
	In this appendix, we collect the first four non-vanishing orders of the expansion \eqref{ExpEffAct} in terms of \eqref{Hexp} around the background Galileon metric \eqref{Gbar}. These results in terms of the $H_{k}^{\mu\nu}$ are further used for the subsequent expansion in terms of the Galileon fields $\pi$, performed in App.~\ref{ExpInPi}. Note that integration by parts rules would allow to simplify the results provided in this Appendix already at the level of the $H_k^{\mu\nu}$. However, as explained in App.~\ref{ccheck}, since the main purpose of this expansion is to obtain a non-trivial crosscheck with a previously derived on-shell result in momentum space, the implementation of integration by parts identities is much simpler realized in momentum space. The results for the three-point and four-point function have been obtained with the tensor-algebra bundle \texttt{xAct} \cite{MartinGarcia,Brizuela2009,Nutma2014}.  
	
	\subsection{Two-point function}
	\begin{align}
	\Gamma^{\mathrm{div}}_{1,2}=\frac{1}{2!}\frac{\Lambda}{480}\int_{\mathcal{M}}\mathrm{d}^4x\left[-2 \bar{\partial}^2H^1_{\alpha \beta} \bar{\partial}^2H_1^{\alpha \beta} -  \left(\bar{\partial}^2H_1\right)^2\right].\label{Var2G1}
	\end{align}
	
	\subsection{Three-point function}
	\begin{align}
	\Gamma^{\mathrm{div}}_{1,3}={}&\frac{1}{3!}\frac{\Lambda}{320}\int_{\mathcal{M}}\mathrm{d}^4x\left[
	8 H_1^{\alpha \beta} \bar{\partial}^2H_1^{\alpha\rho} \bar{\partial}^2H_1^{\beta \rho} - 4 \bar{\partial}^2H_1^{\alpha \beta} \bar{\partial}^2H_2^{\alpha \beta} + 2 \bar{\partial}^2H_1^{\alpha \beta} \bar{\partial}^2H_1^{\alpha \beta} H_1\right. \nonumber \\ 
	& - 4 H_1^{\alpha \beta} \bar{\partial}^2H_1^{\alpha \beta} \bar{\partial}^2H_1 -  H_1 \left(\bar{\partial}^2H_1\right)^2 + 2 \bar{\partial}^2H_1 \bar{\partial}^2H_2 \nonumber \\ 
	& + 4 \bar{\partial}^2H_1^{\beta \rho} \partial_{\alpha}H_1^{\beta \rho} \partial^{\alpha}H_1 -  \bar{\partial}^2H_1 \partial_{\alpha}H_1 \partial^{\alpha}H_1 + 12 \bar{\partial}^2H_1^{\alpha \beta} \partial_{\alpha}H_1^{\rho \sigma} \partial_{\beta}H_1^{\rho \sigma} \nonumber \\ 
	& - 8 H_1^{\alpha \beta} \bar{\partial}^2H_1^{\rho \sigma} \partial_{\beta}\partial_{\alpha}H_1^{\rho \sigma} - 4 \bar{\partial}^2H_2^{\alpha \beta} \partial^{\beta}\partial^{\alpha}H_1 + 4 \bar{\partial}^2H_1^{\alpha \beta} H_1 \partial^{\beta}\partial^{\alpha}H_1 \nonumber \\ 
	& + 4 H_1^{\alpha \beta} \bar{\partial}^2H_1 \partial^{\beta}\partial^{\alpha}H_1 + 12 \partial_{\alpha}H_1^{\rho \sigma} \partial_{\beta}H_1^{\rho \sigma} \partial^{\beta}\partial^{\alpha}H_1 + 8 H_1^{\rho \sigma} \partial_{\beta}\partial_{\alpha}H_1^{\rho \sigma} \partial^{\beta}\partial^{\alpha}H_1 \nonumber \\ 
	& + 2 H_1 \partial_{\beta}\partial_{\alpha}H_1 \partial^{\beta}\partial^{\alpha}H_1 - 4 \partial_{\beta}\partial_{\alpha}H_2 \partial^{\beta}\partial^{\alpha}H_1 - 4 H_1^{\alpha \rho} \partial_{\beta}\partial^{\rho}H_1 \partial^{\beta}\partial^{\alpha}H_1 \nonumber \\ 
	& - 4 \bar{\partial}^2H_1^{\alpha \beta} \partial^{\beta}\partial^{\alpha}H_2 - 8 \bar{\partial}^2H_1^{\beta \rho} \partial^{\alpha}H_1 \partial_{\rho}H_1^{\alpha \beta} + 12 \bar{\partial}^2H_1 \partial_{\beta}H_1^{\alpha \rho} \partial^{\rho}H_1^{\alpha \beta} \nonumber \\ 
	& - 10 \bar{\partial}^2H_1 \partial_{\rho}H_1^{\alpha \beta} \partial^{\rho}H_1^{\alpha \beta} - 4 H_1^{\beta \rho} \partial^{\beta}\partial^{\alpha}H_1 \partial^{\rho}\partial_{\alpha}H_1 + 4 \partial_{\alpha}H_1^{\beta \rho} \partial^{\alpha}H_1 \partial^{\rho}\partial^{\beta}H_1 \nonumber \\ 
	& - 4 \partial^{\alpha}H_1 \partial_{\beta}H_1^{\alpha \rho} \partial^{\rho}\partial^{\beta}H_1 - 4 \partial^{\alpha}H_1 \partial_{\rho}H_1^{\alpha \beta} \partial^{\rho}\partial^{\beta}H_1 + 8 H_1^{\alpha \beta} \bar{\partial}^2H_1^{\rho \sigma} \partial_{\sigma}\partial_{\rho}H_1^{\alpha \beta} \nonumber \\ 
	& - 8 H_1^{\rho \sigma} \partial^{\beta}\partial^{\alpha}H_1 \partial_{\sigma}\partial_{\rho}H_1^{\alpha \beta} - 16 \bar{\partial}^2H_1^{\alpha \beta} \partial_{\beta}H_1^{\rho \sigma} \partial^{\sigma}H_1^{\alpha\rho} - 8 \partial_{\beta}H_1^{\rho \sigma} \partial^{\beta}\partial^{\alpha}H_1 \partial^{\sigma}H_1^{\alpha\rho} \nonumber \\ 
	& + 8 \bar{\partial}^2H_1^{\alpha \beta} \partial_{\rho}H_1^{\beta \sigma} \partial^{\sigma}H_1^{\alpha\rho} + 8 \partial^{\beta}\partial^{\alpha}H_1 \partial_{\rho}H_1^{\beta \sigma} \partial^{\sigma}H_1^{\alpha\rho} + 8 \bar{\partial}^2H_1^{\alpha \beta} \partial_{\sigma}H_1^{\beta \rho} \partial^{\sigma}H_1^{\alpha\rho} \nonumber \\ 
	&\left. + 8 \partial^{\beta}\partial^{\alpha}H_1 \partial_{\sigma}H_1^{\beta \rho} \partial^{\sigma}H_1^{\alpha\rho} - 8 \partial_{\alpha}H_1^{\rho \sigma} \partial^{\beta}\partial^{\alpha}H_1 \partial^{\sigma}H_1^{\beta\rho}\right].\label{Var3G1}
	\end{align}
	
	\subsection{Four-point function}

	The derivatives in \eqref{Var2G1}-\eqref{Var4G1} are understood to act only upon the object to which they are attached to -- not on the total expression to their right. Indices are raised and lowered with $\bar{G}_{\mu\nu}$ and the contracted derivatives are defined as $\bar{\partial}^2:=\left(\bar{G}^{-1}\right)^{\mu\nu}\partial_{\mu}\partial_{\nu}$. While we have made use of \eqref{DivFreeH} in \eqref{Var2G1}-\eqref{Var4G1}.

	\section{Galileon counterterms up to four-point function }
	\label{ExpInPi}
	Using the results \eqref{Var2G1}-\eqref{Var4G1} for the expansion of \eqref{OneLoopG} and inserting \eqref{Gbar} as well as \eqref{H1}-\eqref{H3}, we express the expansion \eqref{ExpEffAct} up to fourth order in terms of the Euclidean metric $\delta_{\mu\nu}$ and the perturbation of the Galileon field $\delta\pi$.
	The results for the three point and four point function have been obtained with the tensor-algebra bundle \texttt{xAct} \cite{MartinGarcia,Brizuela2009,Nutma2014}.  
	
	\subsection{Two-point function}
	\begin{align}
	\Gamma^{\mathrm{div}}_{1,2}[\pi]={}&-\frac{1}{2!}\frac{9}{2}\frac{\Lambda c_3^2}{ M^6}\int_{\mathcal{M}}\mathrm{d}^4x\pi\left(\partial^8\pi\right).\label{TwoPointPi}
	\end{align}
	
	\subsection{Three-point function}
	\begin{align}
	\Gamma^{\mathrm{div}}_{1,3}[\pi]=\frac{1}{3!}\frac{\Lambda}{160 M^9}\int_{\mathcal{M}}\mathrm{d}^4x\Big[&
	27 (95 c_{3}{}^3 + 24 c_{3}{} c_{4}{}) (\partial^{2}{}\pi) \
	(\partial^{4}{}\pi)^2\nonumber\\
	& - 18 (309 c_{3}{}^3 + 20 c_{3}{} c_{4}{}) \
	(\partial^{2}{}\pi)^2 (\partial^{6}{}\pi) \nonumber \\ 
	& + 27 (159 c_{3}{}^3 - 4 c_{3}{} c_{4}{}) (\partial^{4}{}\pi) \
	(\partial^{6}{}\pi) \pi\nonumber\\
	& + 9 (417 c_{3}{}^3 + 28 c_{3}{} c_{4}{}) \
	(\partial^{2}{}\pi) (\partial^{8}{}\pi) \pi 
	\Big].\label{3PGal}
	\end{align}
	In order to arrive at the final form \eqref{3PGal}, we used the integration by parts reduction rules
	\begin{align}
	\pi(\partial_{\mu}\partial_{\nu}\partial_{\rho}\partial^2\pi)(\partial^{\mu}\partial^{\nu}\partial^{\rho}\partial^2\pi)={}&\frac{1}{2}(\partial^2\pi)(\partial_{\mu}\partial_{\nu}\partial^2\pi)(\partial^{\mu}\partial^{\nu}\partial^2\pi)-\pi(\partial_{\mu}\partial_{\nu}\partial^2\pi)(\partial^{\mu}\partial^{\nu}\partial^4\pi),\label{IBP3p1}\\
	\pi(\partial_{\mu}\partial_{\nu}\partial^4\pi)(\partial^{\mu}\partial^{\nu}\partial^2\pi)={}&\frac{1}{2}\left[(\partial^2\pi)(\partial_{\mu}\partial^4\pi)(\partial^{\mu}\partial^2\pi)-\pi(\partial_{\mu}\partial^4\pi)(\partial^{\mu}\partial^4\pi)\right.\nonumber\\
	&\left.\quad-\pi(\partial_{\mu}\partial^6\pi)(\partial^{\mu}\partial^2\pi)\right],\\
	(\partial^2\pi)(\partial_{\mu}\partial^4\pi)(\partial^{\mu}\partial^2\pi)={}&-\frac{1}{2}(\partial^6\pi)(\partial^2\pi)^2,\\
	(\partial^2\pi)(\partial_{\mu}\partial_{\nu}\partial^2\pi)(\partial^{\mu}\partial^{\nu}\partial^2\pi)={}&-(\partial_{\mu}\partial^2\pi)(\partial_{\nu}\partial^2\pi)(\partial^{\mu}\partial^{\nu}\partial^2\pi)-(\partial^2\pi)(\partial_{\mu}\partial^4\pi)(\partial^{\mu}\partial^2\pi),\\
	(\partial_{\mu}\partial^2\pi)(\partial_{\nu}\partial^2\pi)(\partial^{\mu}\partial^{\nu}\partial^2\pi)={}&-\frac{1}{2}(\partial^4\pi)(\partial_{\mu}\partial^2\pi)(\partial^{\mu}\partial^2\pi),\\
	(\partial^4\pi)(\partial_{\mu}\partial^2\pi)(\partial^{\mu}\partial^2\pi)={}&\frac{1}{2}(\partial^6\pi)(\partial^2\pi)^2-(\partial^4\pi)^2(\partial^2\pi),\\
	\pi(\partial_{\mu}\partial^6\pi)(\partial^{\mu}\partial^2\pi)={}&\frac{1}{2}\left[(\partial^2\pi)^2(\partial^6\pi)-\pi(\partial^6\pi)(\partial^4\pi)-\pi(\partial^8\pi)(\partial^2\pi)\right],\\
	\pi(\partial_{\mu}\partial^4\pi)(\partial^{\mu}\partial^4\pi)={}&-\pi(\partial^6\pi)(\partial^4\pi)+\frac{1}{2}(\partial^2\pi)(\partial^4\pi)^2.\label{IBP3p7}
	\end{align}
	
	\subsection{Four-point function}

	The result for the divergent one-loop off-shell contribution to the four-point function could be drastically reduced by deriving similar integration by parts identities as for the three-point function in \eqref{IBP3p1}-\eqref{IBP3p7}. 
	However, since our main intention in deriving this expression is a check of our result for the divergent part of the one-loop effective action in terms of the geometrical formulation \eqref{OneLoopG} and its role as generating functional for the $n$-point one-loop counterterms \eqref{npoint}, the off-shell result \eqref{4PGal} only provides an intermediate result, which is further reduced to the on-shell result in momentum space in Sec. \ref{Crosscheck4p}. In momentum space, integration by parts identities correspond to the simple algebraic identities that follow from momentum conservation. The momentum on-shell result can be checked against the results of a previously performed diagrammatic calculation in \cite{Kampf2014,Heisenberg2019}. 
	
	\section{Crosschecks}
	\label{ccheck}
	We perform chrosschecks of several limiting cases with results known in the literature
	\subsection{Off-shell two-point function}
	A non-trivial crosscheck of \eqref{OneLoopG} is the result obtained in \cite{PaulaNetto2012} for the terms quadratic in $\pi$.
	In \cite{PaulaNetto2012}, the relevant operator $\tilde{\mathcal{L}}_3$ in the Galileon Lagrangian (Lorentzian signature $\Box:=\partial_{\mu}\partial^{\mu}$) with coupling constant $\tilde{c}_{3}$ reads
	\begin{align}
	\tilde{\mathcal{L}}_{3}=\tilde{c}_3\left(\partial\pi\right)^2\Box\pi\,.\label{ShapL3}
	\end{align}
	In contrast, our operator $\mathcal{L}_3$ (Euclidean signature $\Delta:=-\partial_{\mu}\partial^{\mu}$) in \eqref{Gact} with coupling constant $c_3$ reads after integration by parts
	\begin{align}
	\mathcal{L}_{3}={}&\frac{c_3}{M^3}\pi\epsilon^{\mu\nu\rho\sigma}\tensor{\epsilon}{^{\alpha\beta}_{\rho\sigma}}\pi_{\mu\alpha}\pi_{\nu\beta}=-3\frac{c_{3}}{M^3}\left(\partial\pi\right)^2\Delta\pi,\label{weL3}
	\end{align}
	Comparison of \eqref{ShapL3} and \eqref{weL3} implies the identification
	\begin{align}
	\tilde{c}_{3}=-3\frac{c_{3}}{M^3}.\label{c3s}
	\end{align}
	Moreover, we have defined the pole in dimension as $\varepsilon=(4-d)/2$ while the authors of \cite{PaulaNetto2012} have defined the pole in dimension as $\tilde{\varepsilon}=16\pi^2(d-4)$, which implies
	\begin{align}
	\tilde{\varepsilon}=-32\pi^2\varepsilon=-\frac{2}{\Lambda}.\label{epsid}
	\end{align}
	The result for the one-loop counterterm of the two-point function  in Lorentzian signature obtained in equation $(18)$ of \cite{PaulaNetto2012} is
	\begin{align}
	\Gamma_{1,2}^{\mathrm{div}}=-\frac{1}{2}\frac{\tilde{c}_3^2}{\tilde{\varepsilon}}\int_{\mathcal{M}}\mathrm{d}^4 x\,\pi\Box^4\pi\,.\label{Shap2}
	\end{align}
	Inserting \eqref{c3s} and \eqref{epsid} into \eqref{Shap2} and changing to Euclidean signature, we obtain 
	\begin{align}
	\Gamma_{1,2}^{\mathrm{div}}=-\frac{9}{4}\frac{\Lambda c_3^2}{ M^6}\int_{\mathcal{M}}\mathrm{d}^4 x\,\pi\Delta^4\pi,\label{we2}
	\end{align}
	in agreement with \eqref{TwoPointPi} upon identification of $\delta\pi$ with $\pi$. This provides a non-trivial cross check for our general result \eqref{OneLoopG}. 
	
	\subsection{On-shell four-point function}
	We check the final result  \eqref{OneLoopG} for the geometrically defined divergent part of the one-loop effective action and its role as generating functional for all one-loop $n$-point counterterms \eqref{ExpEffAct}, by comparing the on-shell reduction of the result for the four-point counterterm \eqref{4PGal} obtained from the expansion \eqref{ExpEffAct} with the result of the on-shell four-point divergences obtained in a previous Feynman diagrammatic calculation in momentum space \cite{Kampf2014, Heisenberg2019}. Starting from \eqref{4PGal}, we perform the fourth functional derivative with respect to the $\pi_{i}=\delta\pi(x_i)$ around $\pi=0$ and subsequently transform to Fourier space. This results in an expression involving the ten invariants $(k_{i}\cdot k_{j})$, which can be constructed from the four external momenta $k_{i}^{\mu}$, $i=1,\ldots,4$. Not all of the invariants are independent. Momentum conservation $\sum_{i=1}^4k_{i}^{\mu}=0$ allows to express one external momentum in terms of all the others and thereby reduces the independent invariants from ten to six. The four on-shell conditions $k_{i}^2=(k_{i}\cdot k_{i})=0$ further reduce the independent invariants from six to two. Introducing the Mandelstam variables $s_{ij}=(k_{i}+k_{j})^2$, we represent the result for the divergent part of the momentum space four-point one-loop correlation function in a compact form as power-summed symmetric polynomial in the three (redundant) Mandelstam variables $s_{12}$, $s_{23}$, $s_{31}$,
	\begin{align}
	\langle 1,2,3,4\rangle^{\mathrm{div}}_{\mathrm{on-shell}}
	={}&\frac{243}{80}\frac{\Lambda c_3^4}{ M^{12}}\left(s_{12}^2+s_{23}^2+s_{31}^2\right)^3+\frac{3}{20}\frac{\Lambda c_4^2}{ M^{12}}\left(s_{12}^2+s_{23}^2+s_{31}^2\right)^3\nonumber\\
	&{}+\frac{9}{20}\frac{\Lambda c_3^2c_4}{ M^{12}}\left[20\left(s_{12}^6+s_{23}^6+s_{31}^6\right)-3\left(s_{12}^2+s_{23}^2+s_{31}^2\right)^3\right].\label{Crosscheck4p}
	\end{align} 
	This expression coincides with the result obtained in \cite{Kampf2014,Heisenberg2019} and therefore provides a strong check of the general result \eqref{OneLoopG} and the expansion \eqref{ExpEffAct}.
	\bibliographystyle{JHEP}
	\bibliography{QuantumGalileons}{}
\end{document}